\documentclass[notitlepage, longbibliography]{revtex4-1}%
\usepackage{placeins}
\usepackage{pbox}
\usepackage{relsize}
\usepackage{graphicx}
\usepackage{amsmath}
\usepackage{amssymb}
\usepackage{MathShorthand}
\usepackage{subfigure}
\usepackage{verbatim}
\usepackage{float}
\usepackage{textcomp}
\usepackage{epstopdf}
\usepackage{amsfonts}
\usepackage[margin=0.75in]{geometry}%
\providecommand{\U}[1]{\protect\rule{.1in}{.1in}}
\newtheorem{theorem}{Theorem}[section]
\newtheorem{lemma}[theorem]{Lemma}

\newtheorem{result}[theorem]{Result}

\begin{document}
\begin{empty}
\title{A ring of spikes}
\author{Theodore Kolokolnikov$^\star$ and Michael Ward$^\dagger$}
\affiliation{$^\star$ Department of Mathematics and Statistics, Dalhousie University,
Halifax, Canada\\
$^\dagger$Department of Mathematics, University of British Columbia, Vancouver, Canada}
\begin{abstract}
For the Schnakenberg model, we consider a highly symmetric configuration of N spikes whose
locations are located at the vertices of a regular N-gon inside either a unit disk or an annulus. We call such
configuration a ring of spikes. The ring radius
is characterized in terms of the modified Green's function. For a disk, we find that a ring
of 9 or more spikes is always unstable with respect to small eigenvalues. Conversely, a ring of
8 or less spikes is stable inside a disk provided that the feed-rate $A$ is sufficiently large. More
generally, for sufficiently high feed-rate, a ring of $N$ spikes can be stabilized provided that the annulus is
thin enough. As $A$ is decreased, we show that the ring is destabilized due to small eigenvalues first, and then due to
large eigenvalues, although both of these thresholds are separated by an asymptotically
small amount. For a ring of 8 spikes inside a disk, the instability appears to be supercritical, and deforms the ring into
a square-like configuration. For less than 8 spikes, this instability is subcritical and results in spike death.
\end{abstract}
\maketitle
\end{empty}

\section{Introduction}

\begin{figure}[ptb]
\includegraphics[width=0.98\textwidth]{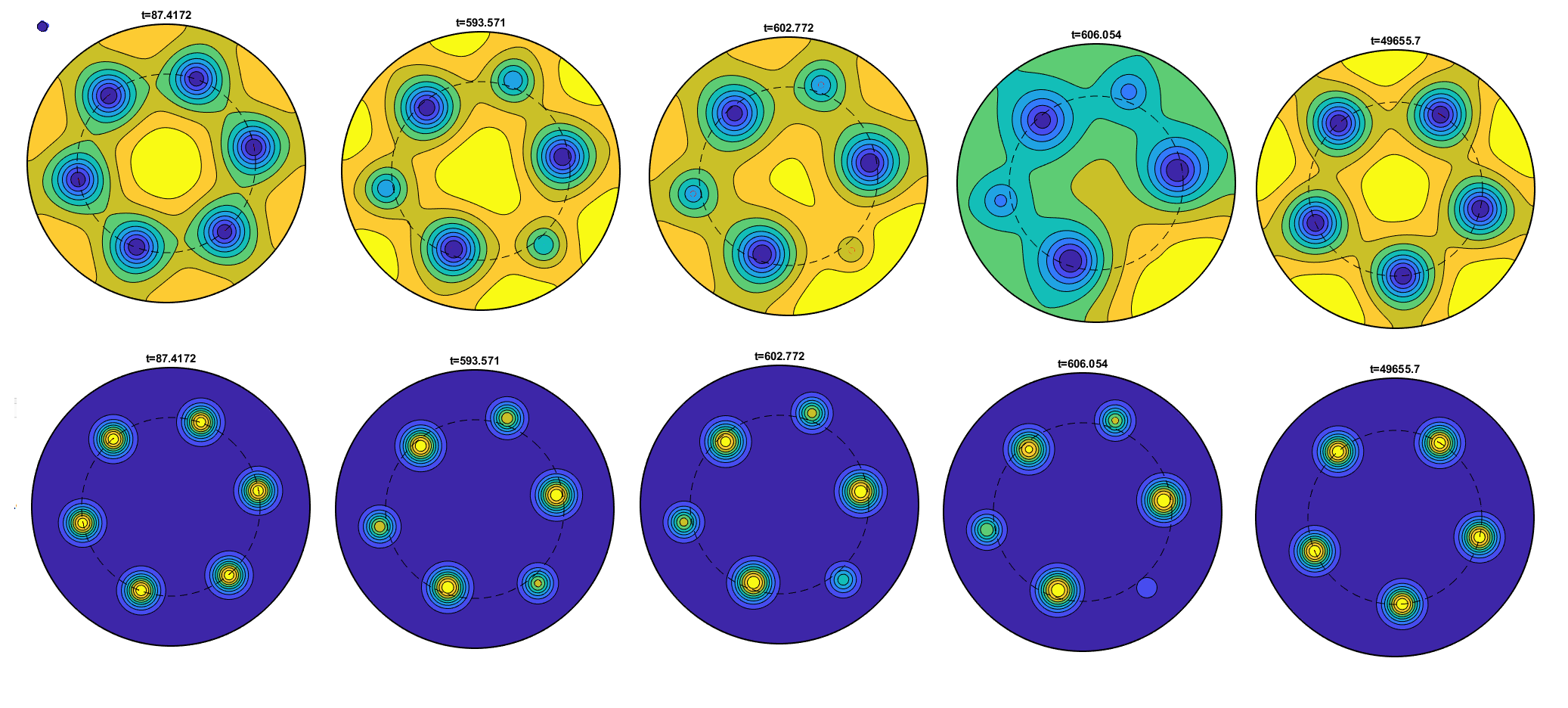}\caption{Transition from a
6-spike to 5-spike ring. Here, $\varepsilon=0.05$ and $A=13.$ The initial
condition was taken to be the equilibrium of a 6-spike state corresponding to
$A=14$. Such a ring is observed to be stable for $A>13.5$ but is unstable for
$A=13.$ Top row is $v(x,t)$ and the bottom row is $u(x,t).$ Half of the spikes
move towards the center and half move towards the boundary, consistent with
the mode $m=3$ small-eigenvalue instability. This eventually leads to the
death of one of the spikes leading to a 5-spike symmetric ring configuration.
Dashed circle shows the theoretical radius of the ring of 6 (or 5, on the last
panel) spikes. We used FlexPDE software to simulate (\ref{pde}).}%
\label{fig:6spikes}%
\end{figure}

The goal of this paper is to study a solution to reaction-diffusion model
consisting of a ring of spikes. This configuration is highly symmetric, which
allows for an in-depth analysis of its stability properties. For simplicity,
we will concentrate on the Schnakenberg model \cite{schnakenberg1979simple}
although similar techniques can be extended to other models. We study the
following version of the Schnakenberg model \cite{xie2017moving}:%
\begin{equation}
u_{t}=\varepsilon^{2}\Delta u-u+u^{2}v,\ \ \ \ 0=\Delta v+A-u^{2}v\frac
{1}{\varepsilon^{2}}\frac{1}{\log\varepsilon^{-1}} \label{pde}%
\end{equation}
with the usual Neumann boundary conditions inside a radially symmetric domain
$\Omega_{b},$ which we take to be either a disk or an annulus of inner radius
$b$ and outer radius 1:%
\begin{equation}
\Omega_{b}=\left\{  x:b<\left\vert x\right\vert <1\right\}  . \label{Omegab}%
\end{equation}
An example of a ring of 6 spikes inside a unit disk is shown in Figure
\ref{fig:6spikes} (left).

The general problem of $N$ spikes in 2D and their stability was considered in
numerous papers. See \cite{wei2008stationary} for a good review and stability
computations for the Schnakenberg model. See also \cite{muratov2001spike,
ward2002existence, wei2003existence, chen2010, kolokolnikov2009spot,
xie2017moving, wong2020spot} for related results in two-dimesions. As is well
known, there are two types of instabilities that are possible:\ due to large
$(O(1)$) or small $(O(\varepsilon^{2})$)\ eigenvalues. Instability triggered
by large eigenvalues induces a \textquotedblleft structural\textquotedblright%
\ or spike profile instability on an O(1) time scale. Numerically, this
instability is observed to be subcritical (see also
\cite{kolokolnikov2021competition, kolokolnikov2020stable} for analysis of
criticality in 1D)\ and quickly leads to a reduction in the number of spikes.
The small-eigenvalue instability induces a spike motion on a slow timescale.
Its criticality depends on the number of spikes as well as domain shape.

In paper \cite{wei2008stationary}, the authors analysed general equilibrium
configurations of $N$ spikes in 2D. They derived a simple threshold on the
feed rate $A$ such that an instability with respect to large eigenvalues is
triggered as $A$ is decreased past that threshold. For for a general spike
equilibrium subject to a natural local-minimality condition related to a
Green's functional, and when $A$ is well above the abovementioned threshold,
they also showed that the small eigenvalues are stable.

However, we will show in this work that this is \emph{not }the case when the
feed rate $A$ is \emph{close} to the large-eigenvalue instability threshold
(within $O\left(  \frac{1}{\log\varepsilon^{-1}}\right)  $ in relative terms).
In fact, as we show in this paper, there is a small-eigenvalue threshold just
above the large eigenvalue threshold which triggers a small-eigenvalue
instability. This instability deforms a ring. In some cases, the deformation
is supercritical, and leads to a nearby non-ring state with the same number of
spikes. In other cases, the deformation is subcritical and leads to a far-away
state and can trigger secondary large-eigenvalue instability, leading to spike
death.

Consider a ring of $N=6$ spikes with $\varepsilon=0.05$; see Figure
\ref{fig:6spikes}. As shown in \S \ref{sec:small} (see Figure \ref{table:A}),
the theory predicts that small eigenvalues are destabilized as $A$ is
decreased below $A_{s}=11.35,$ whereas large ones are destabilized when $A$ is
decreased below $A_{l}=10.18.$ Note that these two thresholds are relatively
close. Numerically, we observe an instability transition at $A\approx13.5$.
The \emph{way }it becomes unstable is shown in Figure \ref{fig:6spikes}. Note
how every second spot around the ring shrinks and moves outwards whereas the
other three spots move inwards, before half of the spots disappear. This
bifurcation appears supercritical. Since $A_{s}$ and $A_{l}$ are very close to
each-other, this deformation eventually triggers a dynamical instability that
leads to eventual destruction of one of the spikes. We remark that in
\cite{wei2008stationary}, the authors computed an instability threshold of
$A_{l,0}=8.89$ (see formula (\ref{Al0})) which is 34\% lower than numerics
indicate. Our prediction of $A_{s}=11.35$ is more accurate (a difference of
about 15\% from $A_{\text{numeric}}\approx13.5$).

\begin{figure}[ptb]%
\[%
\begin{tabular}
[c]{|l|l|l|l|l|l|l|l|l|}\hline
& \multicolumn{2}{|l|}{Arbitrary $\varepsilon$ with $\nu=1/\log\left(
1/\varepsilon\right)  $} & \multicolumn{3}{|l|}{$\varepsilon=0.02$} &
\multicolumn{3}{|l|}{$\varepsilon=0.05$}\\\hline
$N$ & $A_{s}$ & $A_{l}$ & $A_{s}$ & $A_{l}$ & $A_{l,0}$ & $A_{s}$ & $A_{l}$ &
$A_{l,0}$\\\hline
2 & $8.884\nu\left\{  1-0.565\nu\right\}  ^{-1/2}$ & $8.884\nu\left\{
1+0.320\nu\right\}  ^{-1/2}$ & 2.455 & 2.183 & 2.271 & 3.293 & 2.818 &
2.965\\\hline
3 & $13.327\nu\left\{  1-0.165\nu\right\}  ^{-1/2}$ & $13.327\nu\left\{
1+0.484\nu\right\}  ^{-1/2}$ & 3.481 & 3.213 & 3.406 & 4.577 & 4.127 &
4.448\\\hline
4 & $17.769\nu\left\{  1-0.726\nu\right\}  ^{-1/2}$ & $17.769\nu\left\{
1-0.255\nu\right\}  ^{-1/2}$ & 5.033 & 4.698 & 4.542 & 6.814 & 6.201 &
5.931\\\hline
5 & $22.212\nu\left\{  1-0.814\nu\right\}  ^{-1/2}$ & $22.212\nu\left\{
1-0.364\nu\right\}  ^{-1/2}$ & 6.379 & 5.962 & 5.677 & 8.687 & 7.910 &
7.414\\\hline
6 & $26.654\nu\left\{  1-1.157\nu\right\}  ^{-1/2}$ & $26.654\nu\left\{
1-0.709\nu\right\}  ^{-1/2}$ & 8.119 & 7.530 & 6.813 & 11.35 & 10.18 &
8.897\\\hline
7 & $31.096\nu\left\{  1-1.397\nu\right\}  ^{-1/2}$ & $31.096\nu\left\{
1-0.823\nu\right\}  ^{-1/2}$ & 9.912 & 8.946 & 7.949 & 14.20 & 12.18 &
10.38\\\hline
8 & $39.981\nu\left\{  1-3.796\nu\right\}  ^{-1/2}$ & $39.981\nu\left\{
1-1.035\nu\right\}  ^{-1/2}$ & 52.90 & 10.59 & 9.084 & N/A & 14.66 &
11.86\\\hline
\end{tabular}
\]
\caption{Stability thresholds for an $N-$ring inside a unit disk. The ring is
stable when $A>A_{s}$. Note that small-eigenvalue threshold $A_{s}$ is
triggered before the large threshold $A_{l}$, as $A$ is decreased.}%
\label{table:A}%
\end{figure}

\begin{figure}[ptb]
\includegraphics[width=0.55\textwidth]{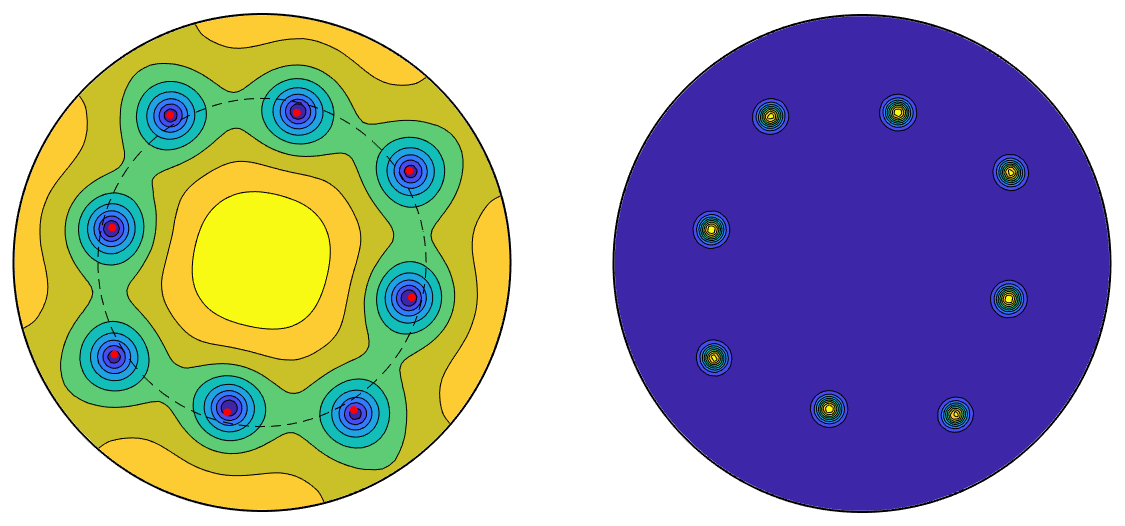}\caption{\textquotedblleft
Square\textquotedblright-type equilibrium with 8 spikes. Here, $A=16.7$ and
$\varepsilon=0.02.$ Dashed line indicates the radius of an 8-spike ring
equilibrium. The 8-spike ring equilibrium is supercritically unstable,
resulting in a nearby square-like stable configuration. Red dots show the
equilibrium of the \emph{reduced} system (\ref{ode}), computed by solving
(\ref{ode}) forward in time until it converged to its equilibrium. The spike
centers of the computed PDE equilibrium were used as initial conditions for
the reduced system (\ref{ode}).}%
\label{fig:8ring}%
\end{figure}

By contrast, Figure \ref{fig:8ring} shows a near-ring steady state of $N=8$
spikes. As we will see in \S \ref{sec:small}, in the \emph{theoretical} limit
$\varepsilon\rightarrow0$ and with $A$ sufficiently big, an 8-ring of spikes
can be stable. However in \emph{practice}, to stabilize such a ring,
$\varepsilon$ needs to be taken too small to have accurate numerical 2D
simulations (smaller than e.g. 0.01). With $\varepsilon=0.02$ our theory
predicts $A_{l}=9.08$ and $A_{s}=52.90$ (c.f. Figure \ref{table:A}). But
self-replication is observed above $A\approx34.78$ (see equation (\ref{Ar})
for a general formula), so we cannot take $A>52.9$ and still retain 8 spikes,
since self-replication will result in more than 8 spots. In Figure
\ref{fig:8ring}, we took $A=16.7$. The result is a \emph{deformed} ring of 8
spikes. In contrast to the 6-ring case, the deformation of an 8-ring appears
to be \emph{supercritical}, and leads to an 8-spike \textquotedblleft
square-type\textquotedblright\ configuration \ as shown in the figure.

\begin{figure}[ptb]
$\includegraphics[width=0.96\textwidth]{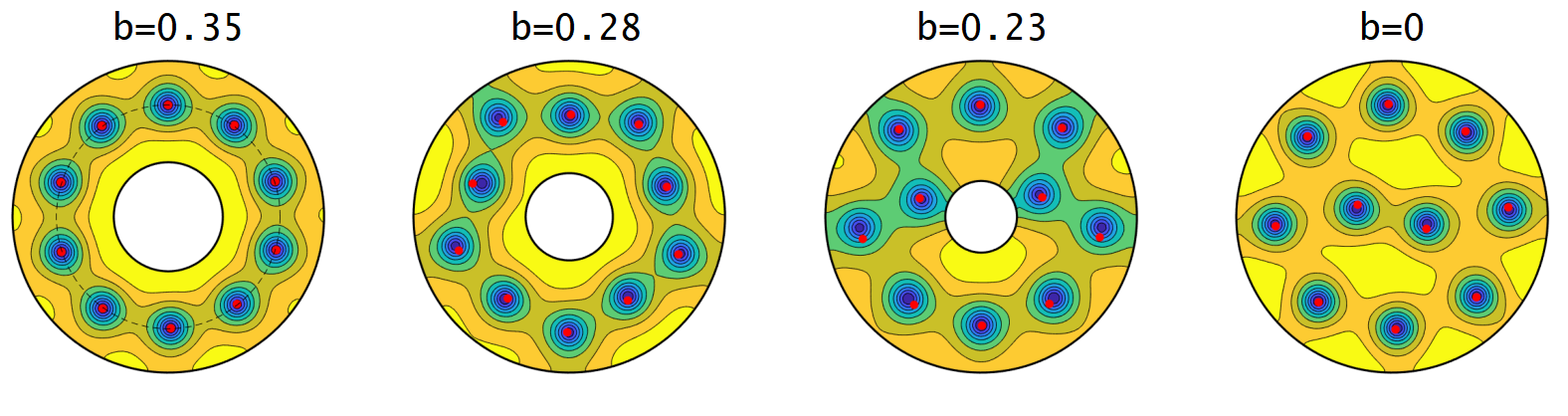}$\caption{Effect of annulus
thickness on ring stability. Here, $A=30$ and $\varepsilon=0.02;$ the $v$
component is shown for several values of inner radius $b$. Each panel shows a
stable equilibrium state computed numerically by solving (\ref{pde}) using
FlexPDE. Red dots show the equilibrium of the reduced system (\ref{ode}),
computed as in Figure \ref{fig:8ring}.}%
\label{fig:10ring}%
\end{figure}

For sufficiently large $A$, namely $A\gg O(\frac{N}{\log\varepsilon^{-1}}),$
it was shown in \cite{wei2008stationary} that large eigenvalues are stable. In
that case, the stability of \emph{small }eigenvalues depends only on the
number of spikes $N$ and the inner radius $b$ of annulus (assuming outer
radius is 1). The following table gives the threshold value of $b_{c}(N)$ such
that $N$ spikes are stable when $b>b_{c}(N):$%
\begin{equation}%
\begin{tabular}
[c]{|l|l|l|l|l|l|l|l|l|l|l|l|l|l|}\hline
$N$ & $\leq8$ & 9 & 10 & 11 & 12 & 13 & 14 & 15 & 16 & 17 & 18 & 19 &
20\\\hline
$b_{c}(N)$ & 0 & 0.174 & 0.293 & 0.356 & 0.412 & 0.450 & 0.488 & 0.516 &
0.545 & 0.567 & 0.589 & 0.607 & 0.625\\\hline
\end{tabular}
\ \label{bc}%
\end{equation}
Figure \ref{fig:10ring} shows a stable 10-spike ring configuration inside an
annulus. Our analysis shows that a 10-spike configuration becomes unstable as
$b$ is decreased below $b=0.293$ (see the table above) Indeed, the ring is
observed to be stable for $b=0.35$ but unstable when $b=0.28.$ The instability
is supercritical when $b$ is close to the threshold value and results in a
zigzag-type configuration near the ring equilibrium radius.



We summarize this paper as follows. In section \ref{sec:r} we characterize the
ring equilibrium radius, and more generally derive the reduced dynamics for
$N$ spikes. This computation is relatively standard; see e.g.
\cite{wei2008stationary, wong2020spot, kolokolnikov2020hexagonal,
kolokolnikov2003reduced}. In \S \ref{sec:large} we compute the stability with
respect to large eigenvalues, specializing to the case of a spike ring. In
\S \ref{sec:small} we linearize the reduced equations of motion to
characterize the stability of a ring with respect to small eigenvalues. An
important aspect of this paper are explicit computations with the Green's
functions and related functional for a disk or an annulus. These are performed
in appendices. We conclude with some open problems in the \S \ref{sec:discuss}.

\section{Equations of spike motion and ring radius\label{sec:r}}

In this section we derive the equilibium ring configuration of $N$ spikes, as
well as reduced equations for spike dynamics. This is a relatively standard
computation, see for example \cite{wei2008stationary, wong2020spot,
ward2002dynamics, kolokolnikov2020hexagonal, kolokolnikov2003reduced}; here,
we follow \cite{kolokolnikov2020hexagonal}. The ring of spikes is an
equilibrium configuration for the reduced dynamics.

We start by deriving equations for reduced spike dynamics; these will
subsequently be used to compute the ring radius and its stability with respect
to small eigenvalues.

\textbf{Inner region. }We will assume that the spike centers $x_{k}$ move on a
slow timescale of $O(\varepsilon^{2})$. This assumption will be seen to be
self-consistent with asymptotic expansions below. As such, we start by
expanding in the inner region near $k$-th spike
\begin{equation}
y=\frac{x-x_{k}(\varepsilon^{2}t)}{\varepsilon}.
\end{equation}
Up to $O(\varepsilon^{2})$ terms, we expand:
\begin{align*}
u  &  =U_{0}(y)+\varepsilon U_{1}(y)+O(\varepsilon^{2});\\
v  &  =V_{0}(y)+\varepsilon V_{1}(y)+O(\varepsilon^{2}).
\end{align*}
The equations for $U_{0},$ and $V_{0}$ become%
\begin{align*}
0  &  =\Delta_{y}U_{0}-U_{0}+U_{0}^{2}V_{0}\\
0  &  =\Delta_{y}V_{0}-\frac{1}{\log\varepsilon^{-1}}U_{0}^{2}V_{0}%
\end{align*}
Next we expand in $\frac{1}{\log\varepsilon^{-1}}.$ Since we only need the
leading order term, to leading order we have $\Delta_{y}V_{0}\sim0$, so we
approximate $V_{0}$ by a constant:
\[
V_{0}\sim v_{k}=v(x_{k}).
\]
The solution for $U_{0}$ is then given by%
\[
U_{0}(y)\sim\frac{w(y)}{v_{k}}%
\]
where $w$ is the ground state satisfying
\begin{equation}
\Delta w-w+w^{2}=0;\ \ \ w\text{ is radially symmetric; \ }w(y)\rightarrow
0\text{ as }\left\vert y\right\vert \rightarrow\infty.
\end{equation}

The leading-order equations (in $\frac{1}{\log\varepsilon^{-1}}$) for
$U_{1},V_{1}$ then become%
\begin{align}
-\frac{x_{k}^{\prime}(s)}{v_{k}}\nabla w  &  =\Delta U_{1}-U_{1}+2wU_{1}%
+w^{2}\frac{V_{1}}{v_{k}^{2}},\label{1514}\\
\Delta V_{1}  &  =0.
\end{align}
Multipying (\ref{1514})\ by $\nabla w$ and integrating, we then obtain the
equation for $x_{k}^{\prime}(s):$%
\begin{equation}
-x_{k}^{\prime}(s)\int\left\vert \nabla w\right\vert ^{2}=-\frac{1}{3v_{k}%
}\int w^{3}\nabla V_{1} \label{1515}%
\end{equation}

\textbf{Outer region. }To estimate the right hand side in (\ref{1515}), we
compute the behaviour of $v$ in the outer region away from spike center. We
estimate%
\[
v(x)\sim T-\sum_{j=1}^{N}S_{j}G(x,x_{j})
\]
where $G(x,x_{j})$ is the Green's function satisfying
\begin{equation}
\left\{
\begin{array}
[c]{l}%
\Delta G-\frac{1}{\pi}=-\delta(x-\xi),\ \ x,\xi\in\Omega_{b},\\
\partial_{n}G=0,\ \ x\in\partial\Omega_{b},\\
\int_{\Omega_{b}}G(x,\xi)dx=0
\end{array}
\right.  \label{G}%
\end{equation}
and $T,\ S_{j}$ satisfy
\begin{align*}
\frac{1}{v_{j}}\frac{1}{\log\varepsilon^{-1}}\int w^{2}dy  &  =S_{j};\\
\sum S_{j}  &  =\left\vert \Omega\right\vert A.
\end{align*}
Recall that the Green's function has the singularity structure,%
\[
G(x,x_{j})=-\frac{1}{2\pi}\log\left\vert x-x_{j}\right\vert +H\left(
x,x_{j}\right)  .
\]
Expanding the outer solution $v$ in the inner variables we then obtain an
expansion%
\[
v(x_{k}+\varepsilon y)\sim T+S_{k}\frac{1}{2\pi}\log\left\vert y\right\vert
-\sum_{j}S_{j}G_{kj}-\varepsilon y\cdot\sum_{j}S_{j}\nabla G_{kj}%
+O(\varepsilon^{2})
\]
where
\begin{equation}
G_{kj}=\left\{
\begin{array}
[c]{c}%
G(x_{k},x_{j}),\ \ \ \ \text{if }k\neq j\\
\frac{1}{2\pi}\log\varepsilon^{-1}+H(x_{j},x_{j}),\text{ \ if }k=j
\end{array}
\right.  ,\ \ \ \ \ \nabla G_{kj}=\left\{
\begin{array}
[c]{c}%
\nabla_{x_{k}}G(x_{k},x_{j}),\ \ \ \ \text{if }k\neq j\\
\nabla_{x}H(x,\xi)|_{\substack{x=x_{k}\\\xi=x_{j}}},\text{ \ if }k=j
\end{array}
\right.  \label{Gkj}%
\end{equation}

Matching with the inner expansion,%
\begin{align*}
\nabla V_{10}  &  \sim-\sum_{j}S_{j}\nabla G_{kj};\\
v_{k}  &  =T-\sum_{j}S_{j}G_{kj}%
\end{align*}
Finally use the following identities identities, see for e.g.
\cite{ward2002dynamics}:%
\begin{equation}
\frac{\int_{%
\mathbb{R}
^{2}}w^{3}dy}{\int_{%
\mathbb{R}
^{2}}w^{2}dy}=3,\ \ \ \ \frac{\int_{%
\mathbb{R}
^{2}}\left\vert \nabla w\right\vert ^{2}dy}{\int_{%
\mathbb{R}
^{2}}w^{2}(y)dy}=1/2,\ \ \ \int_{%
\mathbb{R}
^{2}}w^{2}dy\approx31.04.
\end{equation}
We summarize the spike dynamics as follows.

\begin{result}
Let $x_{k}$ denote the locations of spike centers. Then $x_{k}$ evolve on a
slow timescale according to the following differential-algebraic
system:\bes\label{ode}
\begin{equation}
\frac{dx_{k}}{dt}\sim-\varepsilon^{2}\log\varepsilon^{-1}\frac{2}{\int w^{2}%
}S_{k}\sum_{j=1}^{N}S_{j}\nabla G_{kj};
\end{equation}%
\begin{equation}
\sum_{j=1}^{N}S_{j}=\left\vert \Omega\right\vert A;\ \ \ \frac{\int w^{2}%
}{S_{k}\log\varepsilon^{-1}}=T-\sum_{j=1}^{N}S_{j}G_{kj}.
\end{equation}
\ees

\end{result}

\textbf{Ring equilibrium. }In the case of a ring equilibrium with all spikes
having identical heigth, we have that $S_{k}=S$ for all $k$, so that%
\[
S=S_{k}\sim\frac{\left\vert \Omega\right\vert A}{N}.
\]
The ring equilibrium has the solution of the form
\[
x_{k}=re^{2\pi ik/N}%
\]
We now define%
\begin{equation}
J(r,R,l)=\left\{
\begin{array}
[c]{c}%
G(r,Re^{i2\pi l/N}),\ \ \ \ \text{if }l\neq0\text{ (mod }N)\\
\frac{1}{2\pi}\log\varepsilon^{-1}+H(r,R),\text{ \ otherwise}%
\end{array}
\right.  \label{J}%
\end{equation}
Then $R$ satisfies%
\begin{equation}
\sum_{k=0}^{N-1}J_{r}(R,R,k)=0. \label{1552}%
\end{equation}
The function $J$ as well as the sum in (\ref{1552})\ is computed using Fourier
series decomposition in polar coordinates (see Appendix A). This yields the
following equation for the ring radius $r:$%
\begin{equation}
R^{2}-\frac{1}{2}+\frac{1}{2N}+\frac{1}{R^{-2N}-1}=0. \label{Rb0}%
\end{equation}
This equation was also derived in \cite{wong2020spot} (equation (2.41)); in
addition, the same equation describes an optimal radius
\cite{kolokolnikov2005optimizing} (equation (4.14)), in the context of
optimizing the fundamental Neumann eigenvalue with $N$ small traps on a ring
inside a unit disk.

It is easy to see that (\ref{Rb0})\ has a unique root $R\in\left(  0,1\right)
.$ The following table shows $R$ as a function of $N:$%
\begin{gather}
\text{Ring radius }R\text{ for a ring of }N\text{ spikes}\nonumber\\%
\begin{array}
[c]{cccccccccc}%
N & 2 & 3 & 4 & 5 & 6 & 7 & 8 & 9 & 10\\
R & 0.4536 & 0.5517 & 0.5985 & 0.6251 & 0.6417 & 0.6527 & 0.6604 & 0.6662 &
0.6706
\end{array}
\end{gather}

More generally, for an annulus $\left\vert x\right\vert \in\left(  b,1\right)
,$ the calculations are relegated to Appendix B. As a result, we obtain the
following expression for $R$ in terms of a rapidly converging series:%
\begin{equation}
0=\frac{R^{2}-b^{2}}{\left(  1-b^{2}\right)  }-\frac{1}{2}+\frac{1}{2N}%
+\sum_{p=0}^{\infty}\left\{  \frac{b^{2Np}}{R^{-2N}-b^{2Np}}-\frac
{b^{2N(p+1)}}{R^{2N}-b^{2N(p+1)}}\right\}  . \label{Rb}%
\end{equation}

\section{Stability of a ring, large eigenvalues\label{sec:large}}

We now study the stability of a ring state with respect to large eigenvalues.
We start by linearizing around the ring steady state as%
\[
u(x,t)=u(x)+\phi e^{\lambda t},\ \ \ v(x,t)=v(x)+\phi e^{\lambda t},\ \ \
\]
to obtain the eigenvalue problem,%

\begin{equation}
\lambda\phi=\varepsilon^{2}\Delta\phi-\phi+2uv\phi+u^{2}\psi,\ \ \ \ \Delta
\psi-\left(  2uv\phi+u^{2}\psi\right)  \frac{1}{\varepsilon^{2}\log
\varepsilon^{-1}}=0. \label{1610}%
\end{equation}
Near each spike location $x_{k}$ we let%
\[
x=x_{k}+\varepsilon y;\ \ \Phi_{k}(y)=\phi(x)\text{ \ and \ }\Psi_{k}%
=\psi\left(  x_{k}\right)  .
\]
Then we obtain the eigenvalue problem%
\begin{equation}
\lambda\Phi_{k}=L_{0}\Phi_{k}+w^{2}\frac{\Psi_{k}}{v_{k}};\ \ \ \ \text{where
}\ \ \ L_{0}\Phi:=\Delta\Phi-\Phi+2w\Phi. \label{1611}%
\end{equation}
We estimate%
\begin{equation}
\Psi_{k}\sim C-\sum_{j}\left(  \int\left(  2w\Phi_{j}+w_{j}^{2}\frac{\Psi_{j}%
}{v_{j}^{2}}\right)  dy\right)  \frac{1}{\log\varepsilon^{-1}}G_{kj}
\label{1612}%
\end{equation}
where $G_{kj}$ is given in (\ref{Gkj}); the constant $C$ is determined by
integrating the equation for $\psi$ in (\ref{1610})\ which results in%
\begin{equation}
\sum_{k}\int\left(  2w\Phi_{k}+w^{2}\frac{\Psi_{k}}{v_{k}^{2}}\right)  dy=0.
\label{1613}%
\end{equation}

Together, equations (\ref{1611}), (\ref{1612}) and (\ref{1613}) constitute an
eigenvalue problem for $\lambda.$ Next, we specialize to the case of a ring
spike state. The problem can be decoupled by introducing a \emph{circulant
anzatz }for the eigenfunction of the form%
\[
\Phi_{j}=z^{j}\Phi(y);\ \ \ \Psi_{j}=z^{j}\Psi;\ \ \ \ \ z:=\exp\left(  2\pi
mi/N\right)  ,\ \ \ m=0\ldots N-1,\ \ \ C=0.
\]
Then (\ref{1611})\ becomes%
\[
\lambda\hat{\Phi}=L_{0}\hat{\Phi}+w^{2}\frac{\Psi}{v_{0}^{2}}.
\]
Here, $v_{k}=v_{0}$ is the common height of all $N$ spikes, and $\Psi$
satisfies%
\begin{equation}
\Psi\sim-\left(  2\int w\Phi+\frac{\Psi}{v_{0}^{2}}\int w^{2}\right)  \frac
{1}{\log\varepsilon^{-1}}\sum_{l=0}^{N-1}z^{l}J. \label{500}%
\end{equation}
Here and below, we abbreviate $J=J(R,R,l).$

We now study two cases separately, depending on whether $m=0$ or $m\neq0.$

\textbf{Case 1. }$m=0.$ Then integrating the equation for $\Psi$ in
(\ref{1610})\ we obtain $\Psi=-\frac{\int2w\Phi}{\int w^{2}}v_{0}^{2}$ and
(\ref{1611})\ becomes%
\begin{equation}
\lambda\Phi=L_{0}\Phi-2w^{2}\frac{\int w\Phi}{\int w^{2}}. \label{m0}%
\end{equation}
This case is covered by Theorem 1.4 of \cite{wei1999single}. For convenience,
we state this theorem as follows.

\emph{Theorem (Wei, Theorem 1.4 of \cite{wei1999single}) Consider the
nonlinear eigenvalue problem}%
\begin{equation}
\lambda\Phi=L_{0}\Phi-\chi w^{2}\frac{\int w\Phi}{\int w^{2}}.\label{NLEP}%
\end{equation}
\emph{Suppose that }$\chi>1.$\emph{ Then this problem is stable, that is,
}$\operatorname{Re}\left(  \lambda\right)  <0$\emph{. Suppose that }$\chi
<1.$\emph{ Then (\ref{NLEP}) admits a positive (i.e. unstable)\ eigenvalue
}$\lambda>0.$\emph{ When }$\chi=1,$\emph{ (\ref{NLEP}) has a zero eigenvalue
}$\lambda=0$\emph{ corresponding to the eigenfunction }$\Phi=w.$

It immediately follows that (\ref{m0}) is stable.

\textbf{Case 2. }$m\neq0.$ Then (\ref{500})\ becomes (\ref{NLEP}) with
\begin{equation}
\chi=\frac{2}{1+\left(  \frac{1}{\log\varepsilon^{-1}}\frac{\int w^{2}}%
{v_{0}^{2}}\sum_{l=0}^{N-1}z^{l}J\right)  ^{-1}}.
\end{equation}
By Wei's Theorem, the critical threshold is given when $\chi=1$, which yields%
\begin{equation}
\frac{1}{\log\varepsilon^{-1}}\sum_{l=0}^{N-1}z^{l}J=\frac{v_{0}^{2}}{\int
w^{2}}.\label{2305}%
\end{equation}
Note that $J(R,R,0)\sim\frac{1}{2\pi}\log\varepsilon^{-1}\gg O(1).$ We
therefore define%
\[
\Upsilon\left(  m\right)  :=\sum_{l=0}^{N-1}z^{l}J;\text{ \ and\ \ \ }%
\tilde{\Upsilon}:=\Upsilon-\frac{1}{2\pi}\log\varepsilon^{-1};
\]
Replacing $v_{0}=\frac{1}{\log\varepsilon^{-1}}\frac{N\int w^{2}}{\left\vert
\Omega\right\vert A}$ in (\ref{2305})\ and solving for $A$ we then obtain the
critical threshold for large eigenvalues $A_{l},$ given as%
\begin{equation}
A_{l,m}=\frac{1}{\log\varepsilon^{-1}}\frac{N}{\left\vert \Omega\right\vert
}\left(  2\pi\int w^{2}\right)  ^{1/2}\left(  1+\frac{2\pi}{\log
\varepsilon^{-1}}\tilde{\Upsilon}(m)\right)  ^{-1/2}.\label{Alm}%
\end{equation}
For values of $N\leq8$ on a unit disk, the table below gives numerical values
for $\tilde{\Upsilon}(m):$
\[%
\begin{tabular}
[c]{|l|l|l|l|l|l|l|l|}\hline
N%
$\backslash$%
m & 1 & 2 & 3 & 4 & 5 & 6 & 7\\\hline
2 & 0.0509 &  &  &  &  &  & \\\hline
3 & 0.0771 & 0.0771 &  &  &  &  & \\\hline
4 & 0.148 & -0.0406 & 0.148 &  &  &  & \\\hline
5 & 0.233 & -0.0579 & -0.0579 & 0.233 &  &  & \\\hline
6 & 0.325 & -0.0495 & -0.1129 & -0.0495 & 0.325 &  & \\\hline
7 & 0.4214 & -0.0301 & -0.131 & -0.131 & -0.0301 & 0.4214 & \\\hline
8 & 0.5207 & -0.00471 & -0.1345 & -0.164 & -0.1345 & -0.00471 & 0.5207\\\hline
\end{tabular}
\ \ \ \ \ \
\]

Note that in all cases, $\tilde{\Upsilon}(m)$ attains a minimum at
$m=\left\lfloor N/2\right\rfloor .$ An explicit formula for $\tilde{\Upsilon
}(m)$ with $N$ even and $m=N/2$ is available; it is given by:%
\[
\tilde{\Upsilon}(N/2)=\frac{1}{2\pi}\ln\left(  \frac{4R}{N}\frac{1+R^{N}%
}{1-R^{N}}\right)  .
\]
We now summarize our findings.

\begin{theorem}
\label{thm:large}Let
\begin{equation}
A_{l}:=\max_{1\leq m\leq N-1}A_{l,m}(m).
\end{equation}
Then a ring of $N$ spikes is stable with respect to large eigenvalues provided
that $A<A_{l}.$ When $\Omega$ is a unit disk and $N$ is even, we have an
explicit formula%
\begin{equation}
A_{l}=\frac{1}{\log\varepsilon^{-1}}\frac{N}{\pi}\left(  2\pi\int
w^{2}\right)  ^{1/2}\left(  1+\frac{1}{\log\varepsilon^{-1}}\ln\left(
\frac{4R}{N}\frac{1+R^{N}}{1-R^{N}}\right)  \right)  ^{-1/2}. \label{Al}%
\end{equation}

\end{theorem}

Note that to leading order, $A_{l}\sim A_{l,m}\sim A_{l0}$ as $\varepsilon
\rightarrow0,$ where%
\begin{equation}
A_{l0}:=\frac{N}{\left\vert \Omega\right\vert }\frac{1}{\log\varepsilon^{-1}%
}\left(  2\pi\int w^{2}\right)  ^{1/2}. \label{Al0}%
\end{equation}
Indeed, this recovers the thresold computed in \cite{wei2008stationary} for an
arbitrary configuration of $N$ spikes. However in practice, the $\log
\varepsilon$ correction makes a significant difference. Consider for example
the case $N=8,$ $\varepsilon=0.05.$ Then formula (\ref{Al0}) yields
$A_{l0}=11.86$ whereas $A_{l}=14.66$, so that $O(1/\log\varepsilon)$ terms
contribute about 25\% increase to the instability threshold.

\section{Small eigenvalues\label{sec:small}}

Small eigenvalues control the motion of the spikes.\ They can be computed by
linearizing the reduced ODE\ (\ref{ode})\ around its steady state. Numerical
experiments indicate that the dominant small-eigenvalue instability of a ring
results in a radial motion:\ half of the spikes move inside and half outside
the ring. Thus, we make a simplifying assumption where $k-$th spike is
restricted to move along a ray $\theta=2\pi k/N$. The restricted problem, up
time-rescaling, becomes:%

\[
r_{k}^{\prime}=-S_{k}\sum_{l=0}^{N-1}S_{k+l}J_{r}(r_{k,}r_{k+l},l)\text{
\ with }\theta_{l}=2\pi l/N
\]
with
\[
\sum S_{k}=\left\vert \Omega\right\vert A,
\]%
\[
\frac{1}{\log\varepsilon^{-1}}\frac{\int w^{2}}{S_{k}}=T-\sum_{l=0}^{N}%
S_{k+l}J(r_{k,}r_{k+l},l).
\]
We now linearize around the equilibrium radius $r_{k}=R$ using circular
Fourier series:%
\[
r_{k}=R+\phi z^{k}e^{\lambda t};\ \ \ \ S_{k}=S+\psi z^{k}e^{\lambda
t};\ \ \ z=\exp\left(  2\pi mi/N\right)  ,\ \ \ m=0\ldots N
\]
We then obtain:%
\begin{align*}
\lambda\phi &  =-\phi S^{2}\sum_{l=0}^{N-1}\left(  J_{rr}+J_{rR}z^{l}\right)
-\psi S\sum_{l=0}^{N-1}z^{l}J_{r},\\
\frac{1}{\log\varepsilon^{-1}}\frac{\int w^{2}}{S^{2}}\psi &  =\sum
_{l=0}^{N-1}\psi z^{l}J+S\left(  J_{r}+J_{R}z^{l}\right)  \phi
\end{align*}
Here and below, $J$ denotes $J(R,R,l)$ as defined in (\ref{J}), and we have
used the fact that $\sum_{l=0}^{N-1}J_{r}=0$ for the equilibrium radius $R.$

Eliminating $\psi$ we obtain a single expression for the eigenvalue
$\lambda:$
\begin{equation}
\frac{\lambda}{S^{2}}=-\sum_{l=0}^{N-1}\left(  J_{rr}+J_{rR}z^{l}\right)
-\frac{1}{\frac{1}{\log\varepsilon^{-1}}\frac{\int w^{2}}{S^{2}}-\sum
_{l=0}^{N-1}z^{l}J}\left(  \sum_{l=0}^{N-1}J_{r}z^{l}\right)  ^{2};
\label{10:00}%
\end{equation}
above we used the fact that $J(r,R,l)=J(R,r,l)$ so that $J_{R}(R,R)=J_{r}%
(R,R)$. Letting $\Upsilon(r,R,m)=\sum_{l=0}^{N-1}z^{l}J(r,R,l),$ we obtain
\[
\frac{\lambda}{S^{2}}=-\Upsilon_{rr}(R,R,0)-\Upsilon_{rR}(R,R,m)-\frac
{\Upsilon_{r}^{2}(R,R,m)}{\frac{\kappa}{S^{2}}-\Upsilon(R,R,m)}%
\]
Recall that
\[
S=\frac{\left\vert \Omega\right\vert A}{N};\ \ \Upsilon=\frac{1}{2\pi}%
\log\varepsilon^{-1}+\tilde{\Upsilon}%
\]
where $\tilde{\Upsilon}$ is independent of $\varepsilon.$ It follows that in
the limit $A\gg O(\frac{1}{\log\varepsilon^{-1}})$, the leading-order
stability of small eigenvalues is determined by the sign of $\Upsilon
_{rr}(R,R,0)+\Upsilon_{rR}(R,R,m).$ This quantity is equivalent to local
minimizer condition of the Green's functional from \cite{wei2008stationary},
specialized to a ring of spikes. In addition, recall that $A_{l}=O(\frac
{1}{\log\varepsilon^{-1}})$ so that $A\gg O(\frac{1}{\log\varepsilon^{-1}})$
automatically implies stability with respect to large eigenvalues. We
summarize this as follows.

\begin{result}
Define%
\[
\Lambda(m):=-\Upsilon_{rr}(R,R,0)-\Upsilon_{rR}(R,R,m).
\]
Suppose that $\Lambda(\left\lfloor N/2\right\rfloor )<0$ and moreover,%
\begin{equation}
A\gg O\left(  \frac{1}{\log\varepsilon^{-1}}\right)  . \label{957}%
\end{equation}
Then the ring of $N$ spikes is stable with respect to both small and large
eigenvalues in the limit (\ref{957}). Conversely, if $\Lambda(\left\lfloor
N/2\right\rfloor )>0$ then the ring is unstable for any $A.$

For a disk domain, $\Lambda(m)$ and $\Lambda(\left\lfloor N/2\right\rfloor )$
are explicitly given by (\ref{Lam(m)}) and (\ref{Lam(N/2)}), respectively.
\end{result}

The following table lists the value of $\Lambda(m)$ on a unit disk, using
formula (\ref{Lam(m)}):%
\[%
\begin{array}
[c]{c}%
\Lambda(m)\\%
\begin{tabular}
[c]{|l|l|l|l|l|l|l|l|l|l|}\hline
$N\backslash m$ & 1 & 2 & 3 & 4 & 5 & 6 & 7 & 8 & 9\\\hline
2 & -0.7545 &  &  &  &  &  &  &  & \\\hline
3 & -0.9955 & -0.9955 &  &  &  &  &  &  & \\\hline
4 & -1.2006 & -0.9851 & -1.2006 &  &  &  &  &  & \\\hline
5 & -1.3886 & -0.9722 & -0.9722 & -1.3886 &  &  &  &  & \\\hline
6 & -1.5753 & -0.9682 & -0.7319 & -0.9682 & -1.5753 &  &  &  & \\\hline
7 & -1.7699 & -0.9795 & -0.5129 & -0.5129 & -0.9795 & -1.7699 &  &  & \\\hline
8 & -1.9750 & -1.0074 & -0.3213 & -0.0825 & -0.3213 & -1.0074 & -1.9750 &  &
\\\hline
9 & -2.1901 & -1.0501 & -0.1548 & \textbf{0.3121} & \textbf{0.3121} &
-0.1548 & -1.0501 & -2.1901 & \\\hline
10 & -2.4130 & -1.1043 & -0.0082 & \textbf{0.6747} & \textbf{0.9036} &
\textbf{0.6747} & -0.0082 & -1.1043 & -2.4130\\\hline
\end{tabular}
\end{array}
\]

It shows that the dominant mode corresponds to $m=\left\lfloor
N/2\right\rfloor ;$ moreover a ring of $N\geq9$ spikes is unstable for any
$A.$

\begin{result}
A spike ring with nine or more spikes is unstable inside a unit disk. A ring
of 8 or less spikes is stable in the limit $A\gg O\left(  \frac{1}%
{\log\varepsilon^{-1}}\right)  .$
\end{result}

Note that condition $\Lambda<0$\ alone \emph{does not} guarantee ring
stability when $A$ is of $O\left(  \frac{1}{\log\varepsilon^{-1}}\right)  .$
The full stability characterisation is obtained by setting $\lambda=0$ in
(\ref{10:00}). Upon substituting $\lambda=0$ and $S=\frac{\left\vert
\Omega\right\vert A}{N}$ in (\ref{10:00}) and solving for $A,$ we obtain the
following small-eigenvalue threshold which exists even when $\Lambda(m)<0$ for
all $m\in\left(  1,N\right)  :$%
\begin{equation}
A_{s,m}=\frac{1}{\log\varepsilon^{-1}}\frac{N}{\left\vert \Omega\right\vert
}\left(  2\pi\int w^{2}\right)  ^{1/2}\left\{  1+\frac{2\pi}{\log
\varepsilon^{-1}}\left[  \tilde{\Upsilon}(R,R,m)+\frac{\Upsilon_{r}%
^{2}(R,R,m)}{\Lambda(m)}\right]  \right\}  ^{-1/2}. \label{Asm}%
\end{equation}
Numerics show that the largest $A_{s,m}$ is attained when $m=\left\lfloor
N/2\right\rfloor .$

Let us now contrast the small-eigenvalue threhsold $A_{s,m}$ in (\ref{Asm})
with the the large-eigenvalue threshold $A_{l,m}$ in (\ref{Alm}). Note that
both $A_{s,m}$ and $A_{l,m}$ converge to $A_{l0}\sim\frac{1}{\log
\varepsilon^{-1}}\frac{N}{\left\vert \Omega\right\vert }\left(  2\pi\int
w^{2}\right)  ^{1/2}$ as $\varepsilon\rightarrow0$, which is independent of
the mode $m.$ Moreover, suppose that $\Lambda(m)<0.$ (i.e. the ring is stable
for sufficiently large $A$). Then $\frac{\Upsilon_{r}^{2}(R,R,m)}{\Lambda
(m)}<0$ and it immediately follows from (\ref{Asm}) and (\ref{Alm}) that
$A_{s}>A_{l}.$ We conclude that that small eigenvalues are destabilized before
the big eigenvalues (although both thresholds agree at leading order). This is
indeed the case whenever a ring is stable for sufficiently large $A$ (so that
$\Lambda<0$). We summarize this as follows.

\begin{result}
Suppose that an $N$-ring is stable for sufficiently large $A.$ Let
$A_{s}=A_{s,\left\lfloor N/2\right\rfloor }.$ Then the ring is stable when
$A>A_{s},$ but becomes unstable with respect to small values as $A$ is
decreased below $A_{s}.$
\end{result}

For a unit disk, this result applies to $N\leq8$, since a ring of 9 or more
spikes is unstable for any $A.$ More generally, (\ref{bc}) gives the radius
$b_{c}(N)$ such that $N$ spikes are stable for large $A$ when $b>b_{c}(N).$
This table is generated by solving $\Lambda(\left\lfloor N/2\right\rfloor )=0$
for $b.$ Any number of spikes can be stabilized for sufficiently thin annulus.
Deriving the exact asymptotics of this stabilization is an open question.

\section{Discussion\label{sec:discuss}}

We have performed the stability analysis of a ring solution inside a unit disk
or an annulus $\Omega_{b}=\left\{  x:b\leq\left\vert x\right\vert
\leq1\right\}  $. We found that there are two distinct mechanisms whereby a
ring can undergo an instability. First, if $A$ is sufficiently large, the ring
can be stabilized by making the annulus sufficiently thin. For a unit disk
($b=0$), the magic number is $N=8:$ less than 9 spikes are stable inside a
disk assuming $A$ is sufficiently large (and $\varepsilon$ sufficiently
small). Conversely, a ring of 9 or more spikes is unstable inside a unit disk
but can be stabilized by increasing $b,$ as shown in (\ref{bc}). In fact, the
thinner the annulus, the more spikes can be stable along the ring. It is an
open question to work out the asymptotics for stability of a ring in the limit
of thin annulus.

\begin{figure}[ptb]
\includegraphics[width=0.95\textwidth]{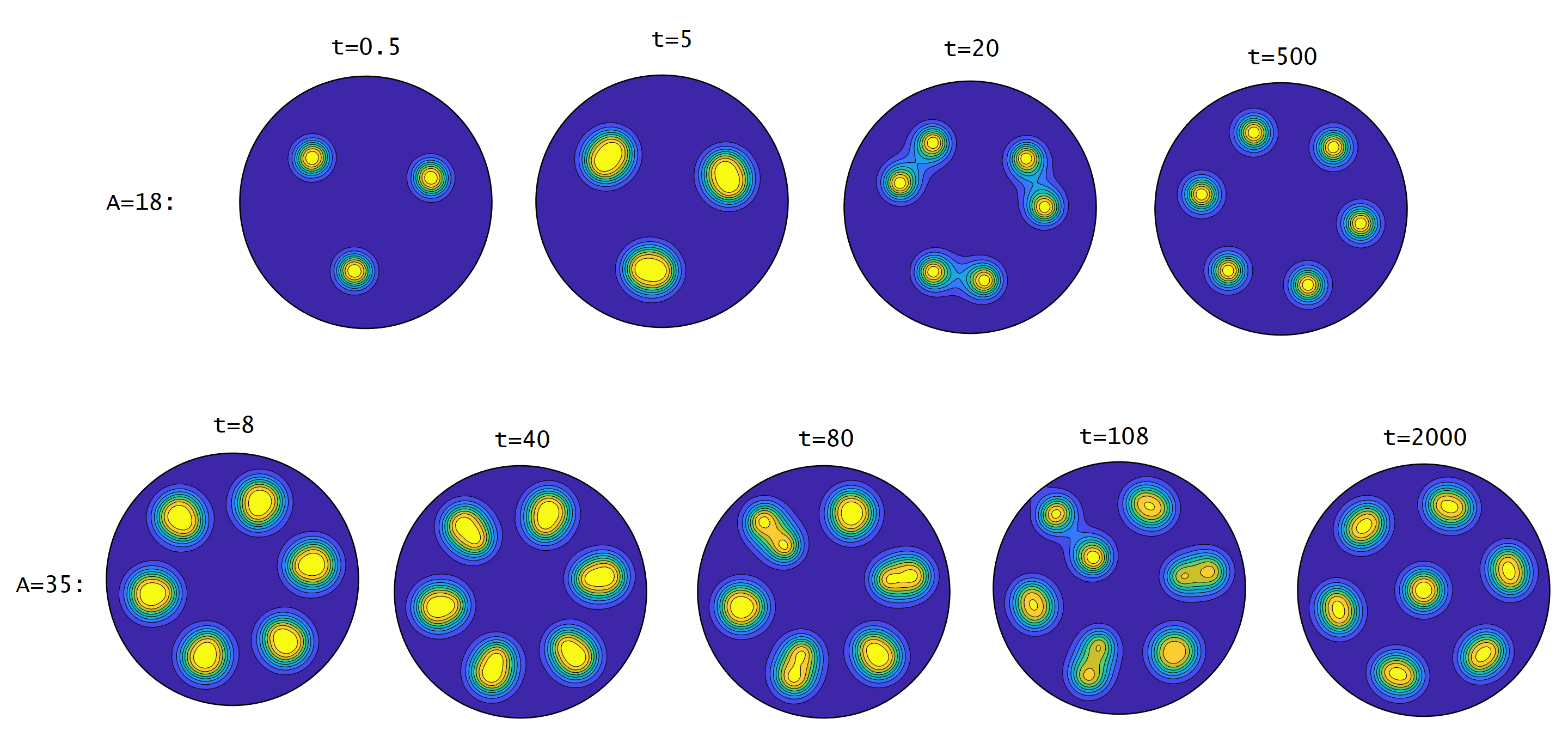}\caption{Self-replication of
an $N$-ring pattern. Here, $\varepsilon=0.05$. Top row:\ $A=15.$ All three
spots split at the same time and the direction of splitting is parallel to the
boundary. Bottom row:\ $A=36$. All six spots undergo an initial deformation
but eventually only one splits. The direction of splitting is perpendicular to
the boundary.}%
\label{fig:split}%
\end{figure}

On the other hand, an $N-$ring can become unstable \emph{regardless} of the
$b$ if $A$ is decreased sufficiently. It was previously known that such an
instability is triggered due to large eigenvalues when $A$ is decreased below
$A_{l0}$ in (\ref{Al0}). We have shown that there is also a
\emph{small-eigenvalue instability} $A_{s}$ just above $A_{l0}$ which triggers
an instability. In particular for an $8-$ring on a disk, this small-eigenvalue
instability explains the square-type pattern of 8 spikes observed (c.f. Figure
\ref{fig:8ring}). Numerics indicate that it is supercritical for an 8-ring on
a unit disk but subcritical for a 6-spike ring. It is an open question to
characterize the criticality analytically.

Another well-known instability for the Schnakenberg model is
spike-replication, which occurs when $A$ is sufficiently increased. Following
the analysis in \cite{kolokolnikov2009spot}, it can be shown that
self-replication of an $N$-ring occurs when $A$ is increased past $A_{r}$,
where%
\begin{equation}
A_{r}=\sqrt{\frac{1}{\log\varepsilon^{-1}}}\frac{N}{\left\vert \Omega
_{b}\right\vert }4.3\cdot2\pi. \label{Ar}%
\end{equation}
Note that unlike competition thresholds $A_{l}$ and $A_{s}$, the formula for
$A_{r}$ is independent of ring radius. This is due to the high symmetry (all
heights being the same) of the ring. Figure \ref{fig:split} illustrates this
phenomenon. Generally, the stability region is $A_{s}<A<A_{r}.$ For an 8-ring,
we have $A_{s}\approx A_{r}\approx33.8$ when $\varepsilon=0.016,$ and the
ordering $A_{s}<A_{r}$ holds as long as $\varepsilon<0.016.$ In particular, no
stable 8-ring can exist if $\varepsilon=0.02$ regardless of the choice of $A$
(c.f. as in figure \ref{fig:8ring}).

Let us conclude with some open questions regarding ring self-replication.
Figure \ref{fig:split} suggests that the \emph{direction} of replication
depends on the particular configuration. In the case of a 3-ring, the
direction of self-replication is parallel to the boundary, whereas in the case
of 6-ring, it is orthogonal to the boundary. Furthermore, \emph{number of
spots that simulateneously self-replicate} also varies with $N.$ For example,
an \textquotedblleft aborted\textquotedblright\ self-replication is observed
in 2nd row of Figure \ref{fig:split}:\ initially, all 6 spots exhibit
self-replication instability; later on, only three of the six spots continue
to replicate, but eventually only one spot succeeds in fully replicating.
Further experiments (not shown)\ indicate that the number of self-replicating
events is very sensitive to how much the feed rate $A$ is above the
self-replication threshold $A_{r}$, as well as the total number of spots.

\section*{Appendix A:\ Green's function on a disk}

In this appendix we summarize the computations involving the Green's function
(\ref{G}) on a unit disk $\Omega_{0}=\left\{  x:\left\vert x\right\vert
<1\right\}  .$

Let $r=\left\vert x\right\vert ,\ \ R=\left\vert \xi\right\vert ,$ and let
$\theta$ be the angle between $x,\xi.$ We decompose into Fourier series
as\ follows:%
\begin{equation}
G=\sum_{n=0}^{\infty}\cos(n\theta)g_{n}(r,R);\ \text{and}\ \ \delta
(x-\xi)=\left(  \frac{1}{2\pi}+\sum_{n=1}^{\infty}\frac{1}{\pi}\cos\left(
n\theta\right)  \right)  \frac{\delta(r-R)}{r} \label{Gfourier}%
\end{equation}
so that%
\begin{align*}
\left(  g_{n}\right)  _{rr}+\frac{1}{r}\left(  g_{n}\right)  _{r}-n^{2}g_{n}
&  =-\frac{1}{\pi R}\delta(r-R),\ \ n\geq1,\\
\left(  g_{0}\right)  _{rr}+\frac{1}{r}\left(  g_{0}\right)  _{r}  &
=-\frac{1}{2\pi R}\delta(r-R),\ \ n=0.
\end{align*}
It is straighforward to verify that%
\begin{align}
g_{n}(r,R)  &  =\frac{1}{2\pi}\frac{1}{n}\left\{
\begin{array}
[c]{c}%
r^{n}\left(  R^{n}+R^{-n}\right)  ,\ \ \ r<R\\
R^{n}\left(  r^{n}+r^{-n}\right)  ,\ \ \ R<r<1
\end{array}
\right.  ,\ \ \ n\geq1,\\
g_{0}(r,R)  &  =\frac{1}{2\pi}\left\{
\begin{array}
[c]{c}%
\frac{r^{2}}{2}+C(R),\ \ \ r<R\\
\ln R-\ln r+\frac{r^{2}}{2}+C(R),\ \ \ R<r<1
\end{array}
\right.  .
\end{align}
Here, $C(R)$ is determined via the integral constraint $\int G=0,$ which
yields%
\begin{equation}
C(R)=\frac{R^{2}}{2}-\log R-\frac{3}{4}.
\end{equation}
Next, we need to compute the regular part $H=G+\frac{1}{2\pi}\log\left\vert
x-\xi\right\vert .$ In what follows, we will assume without loss of generality
that $r<R.$ We have the following expansion of the $\log\left\vert
x-\xi\right\vert :$%
\[
\log\left\vert x-\xi\right\vert =\log R-\sum_{n=1}^{\infty}\cos(n\theta
)\frac{r^{n}R^{-n}}{n},\ \ \ r<R
\]
so that%
\[
2\pi H(r,R,\theta)=\frac{r^{2}}{2}+C(R)+\log R+\sum_{n=1}^{\infty}\cos
(n\theta)\frac{1}{n}r^{n}R^{n}.
\]
We remark that these formulas agree with an explicit expression for Green's
function given in \cite{ward2002dynamics}, namely%
\begin{align}
G(x,\xi)  &  =-\frac{1}{2\pi}\log\left(  \left\vert x-\xi\right\vert \right)
+H(x,\xi);\\
H(x,\xi)  &  =\frac{1}{4\pi}\left[  -\log(\left\vert x\right\vert
^{2}\left\vert \xi\right\vert ^{2}+1-2x\cdot\xi)+\left\vert x\right\vert
^{2}+\left\vert \xi\right\vert ^{2}-\frac{3}{2}\right]  .
\end{align}

\textbf{2. Computing }$\Upsilon(r,R,m).$ Next, we compute%
\begin{equation}
\Upsilon(r,R,m)=\sum_{l=0}^{N-1}J(r,R,l)z^{l},\ \ \ z=\exp(2\pi im/N)
\end{equation}
where $J(r,R,l)$ is defined in (\ref{J}). We obtain,
\begin{equation}
J(r,R,l)=\frac{1}{2\pi}\left\{
\begin{array}
[c]{c}%
-\log\varepsilon^{-1}+\frac{r^{2}}{2}+C(R)+\log R+\sum_{n=1}^{\infty}\frac
{1}{n}r^{n}R^{n},\ \ \ \ \ \ \ \ \ \ \ \ l=0\\
\frac{r^{2}}{2}+C(R)+\sum_{n=1}^{\infty}\cos(2\pi ln/N)\frac{1}{n}\left(
r^{n}R^{n}+r^{n}R^{-n}\right)  ,\ \ \ \ \ \ \ \ \ \ 0<l<N
\end{array}
\right.  .
\end{equation}
Recalling that $\sum_{0}^{N}z^{l}=0,$ this yields:
\begin{equation}
2\pi\tilde{\Upsilon}(r,R,m)=\left\{
\begin{array}
[c]{c}%
\log R-\log(1-Rr)+\rho(rR;m)+\rho(r/R;m),\ \ \ m\in\left(  1,N\right) \\
\log R-\log(1-Rr)+\rho(rR;m)+\rho(r/R;m)+\left(  \frac{r^{2}}{2}+C(R)\right)
N,\ \ \ m=0
\end{array}
\right.  \label{1047}%
\end{equation}
where we defined%
\begin{equation}
\rho(a;m):=\sum_{l=1}^{N-1}\sum_{n=1}^{\infty}\cos\left(  \frac{2\pi nl}%
{N}\right)  e^{2\pi mli/N}\frac{a^{n}}{n}.
\end{equation}
Next we show the following.

\begin{lemma}
\label{lem:rho}We have the following explicit formulas:%
\begin{equation}
a\rho^{\prime}(a;m)=\left\{
\begin{array}
[c]{c}%
\frac{N}{1-a^{N}}\left(  \frac{a^{m}+a^{N-m}}{2}\right)  -\frac{a}%
{1-a},\ \ \ \ m\in(0,N).\\
\frac{N}{1-a^{N}}-N-\frac{a}{1-a},\ \ \ m=0.
\end{array}
\right.  \label{arhop}%
\end{equation}
In addition we have the following identities:%
\begin{align}
\lim_{a\rightarrow1}a\rho^{\prime}(a;m)  &  =\left\{
\begin{array}
[c]{c}%
1/2,\ \ \ \ \ m\in(0,N)\\
1/2-N/2,\ \ \ m=0
\end{array}
\right. \label{656a}\\
\lim_{a\rightarrow1}\left(  a\rho^{\prime}(a;m)\right)  ^{\prime}  &
=\frac{1-N^{2}}{12}+\frac{m}{2}\left(  N-m\right)  \label{656b}%
\end{align}
We also have:%
\[
\rho(a;0)=\ln(1-a)-\ln(1-a^{N}).
\]
When $N$ is even and $m=N/2,$ we have\bes\label{rho(N/2)}%
\begin{align}
\rho(a,N/2)  &  =\ln\left(  \frac{1+a^{N/2}}{1-a^{N/2}}\right)  +\ln
(1-a)\label{652a}\\
\rho(1,N/2)  &  =\ln2+\ln\left(  2/N\right)  =\ln(4/N). \label{652b}%
\end{align}
\ees

\end{lemma}

\textbf{Proof of Lemma \ref{lem:rho}. }

Let $f(a)=a\rho^{\prime}(a;m)=\sum_{l=1}^{N-1}\sum_{n=1}^{\infty}\frac{1}%
{2}\left\{  \exp\left(  \frac{2\pi nl}{N}i\right)  +\exp\left(  -\frac{2\pi
nl}{N}i\right)  \right\}  e^{2\pi mli/N}a^{n}$. We have:%
\begin{align*}
\sum_{l=1}^{N-1}\sum_{n=1}^{\infty}\exp\left(  \frac{2\pi\left(  n-m\right)
l}{N}i\right)  a^{n}  &  =-\sum_{n=1}^{\infty}a^{n}+N\sum_{\substack{n=1\ldots
\infty\ \\n=m(\operatorname{mod}N)}}^{\infty}a^{n}\\
&  =-\frac{a}{1-a}+N\frac{a^{m}}{1-a^{N}}%
\end{align*}
and similarly,
\[
\sum_{l=1}^{N-1}\sum_{n=1}^{\infty}\exp\left(  \frac{2\pi\left(  n+m\right)
l}{N}i\right)  a^{n}=\left\{
\begin{array}
[c]{c}%
-\frac{a}{1-a}+N\frac{a^{N-m}}{1-a^{N}},\ \ m\in\left(  0,N\right) \\
-\frac{a}{1-a}+N\frac{1}{1-a^{N}},\ \ m=0
\end{array}
\right.
\]
This yields (\ref{arhop}). Integrating $\rho^{\prime}(a;N/2)$ yields
(\ref{652a}). Taking limits as $a\rightarrow1^{-}$ yields (\ref{656a},
\ref{656b}, \ref{652b}). $\blacksquare$

The radius of the ring satisfies $\Upsilon_{r}(R,R,0)=0.$ From (\ref{1047}%
)\ and Lemma \ref{lem:rho}\ we compute:%
\begin{equation}
2\pi\Upsilon_{r}(R,R,0)=\frac{1}{R}\frac{R^{2N}N}{1-R^{2N}}+\frac{1}{R}%
\frac{1-N}{2}+RN. \label{834}%
\end{equation}
Setting (\ref{834}) to zero yields (\ref{Rb0}).

\textbf{Computing} $\Lambda=\Upsilon_{rR}(R,R,0)+\Upsilon_{rr}(R,R,0).$ Using
(\ref{1047}) and Lemma \ref{lem:rho}\ we compute%
\begin{align*}
2\pi\Upsilon_{rr}(R,R,0)  &  =NR^{2N-2}\frac{\left(  N-1+R^{2N}\right)
}{\left(  1-R^{2N}\right)  ^{2}}+\frac{1}{R^{2}}\left\{  \frac{1-N^{2}}%
{12}-\frac{1}{2}+\frac{N}{2}\right\}  +N\\
2\pi\Upsilon_{rR}(R,R,m)  &  =\frac{N}{2}R^{2N-2}\frac{\left(  N-m\right)
\left(  R^{2m}+R^{-2m}\right)  +m\left(  R^{2(N-m)}+R^{2\left(  m-N\right)
}\right)  }{\left(  1-R^{2N}\right)  ^{2}}-\frac{1}{R^{2}}\left\{
\frac{1-N^{2}}{12}+\frac{m}{2}\left(  N-m\right)  \right\}
\end{align*}
so that%

\begin{equation}
-2\pi\Lambda=NR^{2N-2}\frac{\left(  N-1+R^{2N}\right)  +\frac{\left(
N-m\right)  }{2}\left(  R^{2m}+R^{-2m}\right)  +\frac{m}{2}\left(
R^{2(N-m)}+R^{2\left(  m-N\right)  }\right)  }{\left(  1-R^{2N}\right)  ^{2}%
}+\frac{1}{2R^{2}}\left\{  -1+N-m\left(  N-m\right)  \right\}  +N.
\label{Lam(m)}%
\end{equation}
In particular, the \textquotedblleft middle\textquotedblright\ mode $m=N/2$
(with $N$ even) yields:%
\begin{equation}
\ 2\pi\Lambda\left(  N/2\right)  =-NR^{2N-2}\frac{\left(  N-1+R^{2N}\right)
+N\left(  R^{N}+R^{-N}\right)  }{\left(  1-R^{2N}\right)  ^{2}}+\frac
{1}{8R^{2}}\left(  N-2\right)  ^{2}-N. \label{Lam(N/2)}%
\end{equation}

\section*{Appendix B:\ Green's function and ring radius in an annulus}

For the annular domain $\Omega_{b}=\left\{  x:b<\left\vert x\right\vert
<1\right\}  $ we decompose in Fourier series as in (\ref{Gfourier}). We then
obtain
\begin{align*}
g_{n}(r,R) &  =\frac{1}{2\pi n\left(  1-b^{2n}\right)  }\left\{
\begin{array}
[c]{c}%
\left(  r^{n}+r^{-n}\right)  \left(  R^{n}+R^{-n}b^{2n}\right)
,\ \ \ \ \ R<r<1\\
\left(  R^{n}+R^{-n}\right)  \left(  r^{n}+r^{-n}b^{2n}\right)  ,\ \ \ b<r<R
\end{array}
\right.  ,\ \ n\geq1\\
g_{0}(r,R) &  =\frac{1}{2\pi(1-b^{2})}\left\{
\begin{array}
[c]{c}%
-b^{2}\ln\left(  r\right)  +\frac{r^{2}}{2}+C(R),\ \ \ \ R_{i}<r<R\\
-\ln\left(  r\right)  +\left(  1-b^{2}\right)  \ln R+\frac{r^{2}}%
{2}+C(R),\ \ \ R<r<1
\end{array}
\right.  ,\ \ n=0
\end{align*}
The constant $C$ is obtained by setting $\int_{b}^{1}g_{0}rdr=0$ yielding%
\[
C=\frac{R^{2}}{2}-\log R-\frac{3}{4}(1+b^{2})+\frac{b^{2}}{b^{2}-1}\log b.
\]

Next we compute the regular part. As before, we need only consider the case
$r<R.$ Write%
\[
H=G+\frac{1}{2\pi}\ln\left\vert x-\xi\right\vert =h_{0}(r,R)+\sum
_{n=1}^{\infty}\cos(n\theta)h_{n}(r,R)
\]
Expanding, for for $r<R,$ we have%
\[
g_{n}=\frac{1}{2\pi n\left(  1-b^{2n}\right)  }\left(  R^{n}r^{n}+R^{-n}%
r^{n}+b^{2n}R^{n}r^{-n}+b^{2n}R^{-n}r^{-n}\right)
\]
so that, for $r<R,$%
\begin{align*}
h_{n}  &  =\frac{1}{2\pi n}\left\{  \frac{R^{n}r^{n}+R^{-n}r^{n}+b^{2n}%
R^{n}r^{-n}+b^{2n}R^{-n}r^{-n}}{\left(  1-b^{2n}\right)  }-r^{n}R^{-n}\right\}
\\
&  =\frac{1}{2\pi n}\left\{  \frac{R^{n}r^{n}+b^{2n}R^{-n}r^{n}+b^{2n}%
R^{n}r^{-n}+b^{2n}R^{-n}r^{-n}}{\left(  1-b^{2n}\right)  }\right\}  ;\\
h_{0}  &  =g_{0}.
\end{align*}
\textbf{Computing the radius. }The radius satisfies $\Upsilon_{r}(R,R)=0.$ We have,%

\[
\Upsilon_{r}=\sum_{l=1}^{N-1}\sum_{n=0}^{\infty}\partial_{r}g_{n}\cos\left(
n2\pi l/N\right)  +\sum_{n=0}^{\infty}\partial_{r}h_{n}%
\]
with%
\begin{align*}
\partial_{r}g_{n}  &  =\frac{R^{n}r^{n}+R^{-n}r^{n}-b^{2n}R^{n}r^{-n}%
-b^{2n}R^{-n}r^{-n}}{2\pi r\left(  1-b^{2n}\right)  },\ \ \ n\geq1,\ \ r<R\\
\partial_{r}h_{n}  &  =\frac{R^{n}r^{n}+b^{2n}R^{-n}r^{n}-b^{2n}R^{n}%
r^{-n}-b^{2n}R^{-n}r^{-n}}{2\pi r\left(  1-b^{2n}\right)  },\ \ n\geq
1,\ \ r<R.
\end{align*}
Define
\begin{equation}
Q(\rho,a)=\sum_{l=1}^{N-1}\sum_{n=1}^{\infty}\frac{\rho^{n}}{1-a^{n}}%
\cos\left(  \frac{2\pi l}{N}n\right)  \text{ \ and \ }P(\rho,a)=\sum
_{n=1}^{\infty}\frac{\rho^{n}}{1-a^{n}}.
\end{equation}
We obtain:%
\begin{align}
2\pi R\Upsilon_{r}(R,R)  &  =Q(R^{2},b^{2})+Q(1,b^{2})-Q(b^{2},b^{2})-Q\left(
\frac{b^{2}}{R^{2}},b^{2}\right) \label{705}\\
&  +P(R^{2},b^{2})-P\left(  \frac{b^{2}}{R^{2}},b^{2}\right)  +\frac
{R^{2}-b^{2}}{1-b^{2}}N.
\end{align}
Next we use the following lemma.

\begin{lemma}
We have%
\begin{align}
P(\rho,a)  &  =\sum_{p=0}^{\infty}\frac{\rho a^{p}}{1-\rho a^{p}%
};\ \ \label{P}\\
Q(\rho,a)+P(\rho,a)  &  =\sum_{p=0}^{\infty}N\frac{\rho^{N}a^{Np}}{1-\rho
^{N}a^{Np}}. \label{Q}%
\end{align}

\end{lemma}

\textbf{Proof.} To show (\ref{P}) we employ a resummation trick as follows:%
\[
\sum_{n=1}^{\infty}\frac{\rho^{n}}{1-a^{n}}=\sum_{n=1}^{\infty}\sum
_{p=0}^{\infty}\rho^{n}a^{np}=\sum_{p=0}^{\infty}\sum_{n=1}^{\infty}\left(
\rho a^{p}\right)  ^{n}=\sum_{p=0}^{\infty}\frac{\rho a^{p}}{1-\rho a^{p}}.
\]
The proof of identity (\ref{Q})\ is similar after writing cosine using complex
exponentials, and is left to the reader. $\blacksquare$

Upon substituting (\ref{P},\ref{Q}), into (\ref{705}) and simplifying, we
obtain (\ref{Rb}).

\bibliographystyle{elsarticle-num}
\bibliography{ver2}

\begin{thebibliography}{10}
\expandafter\ifx\csname url\endcsname\relax
  \def\url#1{\texttt{#1}}\fi
\expandafter\ifx\csname urlprefix\endcsname\relax\def\urlprefix{URL }\fi
\expandafter\ifx\csname href\endcsname\relax
  \def\href#1#2{#2} \def\path#1{#1}\fi

\bibitem{schnakenberg1979simple}
J.~Schnakenberg, Simple chemical reaction systems with limit cycle behaviour,
  Journal of theoretical biology 81~(3) (1979) 389--400.

\bibitem{xie2017moving}
S.~Xie, T.~Kolokolnikov, Moving and jumping spot in a two-dimensional
  reaction--diffusion model, Nonlinearity 30~(4) (2017) 1536.

\bibitem{wei2008stationary}
J.~Wei, M.~Winter, Stationary multiple spots for reaction--diffusion systems,
  Journal of mathematical biology 57~(1) (2008) 53--89.

\bibitem{muratov2001spike}
C.~Muratov, V.~Osipov, Spike autosolitons and pattern formation scenarios in
  the two-dimensional gray-scott model, The European Physical Journal
  B-Condensed Matter and Complex Systems 22~(2) (2001) 213--221.

\bibitem{ward2002existence}
M.~J. Ward, J.~Wei, The existence and stability of asymmetric spike patterns
  for the schnakenberg model, Studies in Applied Mathematics 109~(3) (2002)
  229--264.

\bibitem{wei2003existence}
J.~Wei, M.~Winter, Existence and stability of multiple-spot solutions for the
  gray--scott model in r2, Physica D: Nonlinear Phenomena 176~(3-4) (2003)
  147--180.

\bibitem{chen2010}
W.~Chen, M.~J. Ward, The stability and dynamics of localized spot patterns in
  the two-dimensional Gray-Scott model., SIAM J. Appl. Dynam. Systems, 2010.

\bibitem{kolokolnikov2009spot}
T.~Kolokolnikov, M.~J. Ward, J.~Wei, Spot self-replication and dynamics for the
  schnakenburg model in a two-dimensional domain, Journal of nonlinear science
  19~(1) (2009) 1--56.

\bibitem{wong2020spot}
T.~Wong, M.~J. Ward, Spot patterns in the 2-d schnakenberg model with localized
  heterogeneities, arXiv preprint arXiv:2009.07882 (2020).

\bibitem{kolokolnikov2021competition}
T.~Kolokolnikov, F.~Paquin-Lefebvre, M.~J. Ward, Competition instabilities of
  spike patterns for the 1d gierer--meinhardt and schnakenberg models are
  subcritical, Nonlinearity 34~(1) (2021) 273.

\bibitem{kolokolnikov2020stable}
T.~Kolokolnikov, F.~Paquin-Lefebvre, M.~J. Ward, Stable asymmetric spike
  equilibria for the gierer--meinhardt model with a precursor field, IMA
  Journal of Applied Mathematics 85~(4) (2020) 605--634.

\bibitem{kolokolnikov2020hexagonal}
T.~Kolokolnikov, J.~Wei, Hexagonal spike clusters for some pde's in 2d,
  Discrete \& Continuous Dynamical Systems-B 25~(10) (2020) 4057.

\bibitem{kolokolnikov2003reduced}
T.~Kolokolnikov, M.~J. Ward, Reduced wave green's functions and their effect on
  the dynamics of a spike for the gierer-meinhardt model, European Journal of
  Applied Mathematics 14~(5) (2003) 513--546.

\bibitem{ward2002dynamics}
M.~J. Ward, D.~McInerney, P.~Houston, D.~Gavaghan, P.~Maini, The dynamics and
  pinning of a spike for a reaction-diffusion system, SIAM Journal on Applied
  Mathematics 62~(4) (2002) 1297--1328.

\bibitem{kolokolnikov2005optimizing}
T.~Kolokolnikov, M.~S. Titcombe, M.~J. Ward, Optimizing the fundamental neumann
  eigenvalue for the laplacian in a domain with small traps, European Journal
  of Applied Mathematics 16~(2) (2005) 161--200.

\bibitem{wei1999single}
J.~Wei, On single interior spike solutions of the gierer-meinhardt system:
  uniqueness and spectrum estimates, European Journal of Applied Mathematics
  10~(4) (1999) 353--378.

\end{thebibliography}

\end{document}